\documentclass[12pt]{article}

\usepackage{multirow}
\usepackage{graphicx}

\newcommand{\beq}{\begin{equation}}
\newcommand{\eeq}{\end{equation}}
\newcommand{\bea}{\begin{eqnarray}}
\newcommand{\eea}{\end{eqnarray}}

\begin{document}
\begin{center}
{\Large Gravitationally-Induced Quantum Superpopsition Reduction with Large Extra Dimensions}
\vskip 1cm
J. R. Mureika\\
\vskip .3cm
Department of Physics\\ 
Loyola Marymount University\\
Los Angeles, CA~~90045-8227\\

email: jmureika@lmu.edu
\end{center}

\vskip 2cm

\noindent{\bf Abstract}\\
A gravity-driven mechanism (``objective reduction'') proposed to 
explain quantum state reduction is analyzed in light of the possible 
existence of large extra dimensions in the ADD scenario.  By calculating 
order-of-magnitude estimates for nucleon superpositions, it is shown that 
if the mechanism at question is correct, constraints may be placed on the 
number and size of extra dimensions.  Hence, measurement of superposition
collapse times ({\it e.g.} through diffraction or reflection experiments) could 
represent a new probe of extra dimensions.  The influence of a time-dependent
gravitational constant on the gravity-driven collapse scheme with and
without the presence of extra dimensions is also discussed.


\pagebreak
\section{Introduction}
Since the Copenhagen interpretation of quantum mechanics, the issues of
state vector collapse and the measurement problem have negated a full
mathematical description of the theory.  From the notion of hidden variables or
``incompleteness'' \cite{hv1,hv2,hv3} to the range of 
``Many Worlds''/``enriched
realities'' hypotheses \cite{mw1,mw2,mw3}, the quest for
a dynamical reduction model has been thus far merely academic (see
\cite{collapsemodels} for a comprehensive review of proposed collapse models).

This paper will consider a gravitationally-driven collapse model based on
the work of Di\'osi \cite{diosi} and Penrose \cite{penrose1,penrose2,penrose3} 
in light of
the possible existence of large extra compactified dimensions in the ADD
framework \cite{add}.  This model is shown to behave as a probe of
extra dimensions, in that very distinct collapse signatures can result for
compactification scales of different sizes.  Strict constraints may thus
be placed on the size and number of extra dimensions if the model is taken
to be correct.  
%Alternatively,
%it is shown that should such extra dimensions exist
%and are on the order of a femtometer in size, the gravity-driven model
%suffers serious failures.  

\section{Gravitationally-induced collapse: Objective reduction (OR)}
Gravity has also been pinpointed as a possible cause of superposition
collapse, with the earliest proposal put forth at least as far back as
the mid-1960s through the 1980s \cite{kar1,kar2,kibble}.  Within the 
past 20 years, this 
``philosophy'' has also been addressed by Di\'{o}si \cite{diosi}, 
Ghirardi {\it et al.} \cite{ghirardi}, and Pearle and Squires \cite{pearle}.
Although the framework of Di\'osi and Penrose are distinct, both draw
a similar conclusion concerning the collapse mechanism.  For a review
of four gravity-driven decoherence models including these two, 
the interested reader is directed to \cite{newdiosi}.
Additionally, a resurgence of work involving the 
Schr\"{o}dinger-Newton equations \cite{schronew1,schronew2} has sought to 
further the investigations of gravity's role
in state vector evolution.

The focus of this paper will be on the discussion in Penrose 
\cite{penrose1,penrose2,penrose3}, which like the aforementioned citations
proposes that quantum superpositions should
not just be though of as occurring in the Hilbert space of wavefunctions.
A collapse inevitably occurs because of some interaction with the
environment, and thus it seems reasonable to believe that each possible
outcome (eigenstate) of the superposition is itself continually interacting
with the environment.  Thus, there must be some physical interpretation or
physical correlate of the superposition.  

A novel experiment to measure such a
macroscopic superposition has been proposed, involving the
reflection from a mirror of a single photon which has passed through a 
partially-transmitting filter.  The mirror is thus in such a physical
superposition of distinct states, corresponding to the reflected and transmitted
photon eigenstate \cite{mirror}.  Alternatively, it has recently been
suggested that such a mechanism could serve as a driving mechanism for
cosmogenic neutrino oscillation \cite{neutrino}, hearkening back to the
notion of flavor-dependent violations of the equivalence principle.

Suppose the possible physical correlates are described by wavefunctions 
$|\psi_1\rangle$ and $|\psi_2\rangle$.  Normally the superposition state would
be described by $\alpha |\psi_1\rangle + \beta |\psi_2\rangle$.  However,
if such a physical correlate superposition occurs, then this
inevitably has an impact on the background spacetime geometry, since it seems
likely that the possible outcome states have conformational differences between
one another.  The actual superposition state is described by the entangled
wavefunction $\alpha |\psi_1\rangle |{\cal G}_1\rangle + \beta |\psi_2\rangle
|{\cal G}_2\rangle$, where the states $|{\cal G}_i\rangle$ represents the
spacetime curvature associated with $|\psi_i\rangle$.
This superposition of geometries creates an ill-definition of the time-like Killing vector,
$\partial/\partial t$, since each state will induce slightly different
curvatures.  This curvature instability is ultimately what leads to the 
collapse. 

Following Penrose's prescription in \cite{penrose1},
the difference between curvature configurations is best measured as the
energy difference between them, which can be approximated as the difference
in free-fall vectors integrated over a hypersurface of constant time,

\beq
\Delta_G \approx \int_{\Sigma} (a_1 - a_2)\cdot(a_1 - a_2)\; d^3x
\label{pen1}
\eeq

Noting that $a_k = - \nabla_k^2 \phi = - 4\pi G \rho$, Equation~(\ref{pen1})
can be simplified to
\beq
\Delta_G \sim G \int \int \frac{[\rho_1(x) - \rho_2(x)][\rho_1(y)-\rho_2(y)]}{|x-y|}\; dx~dy
\label{deg}
\eeq
Clearly $\Delta_G$ depends on the density distributions of each ``eigenstate'',
and furthermore the form of Equation~(\ref{deg}) indicates that the important
consideration is the self-energy of the {\it difference} in conformations
of each.  
%In his earlier work, Di\'{o}si concluded that the collapse should
%be mediated by the gravitational interaction energy
\beq
\Delta_G \sim G \int \int \frac{\rho(x)\rho^\prime(y)}{|x-y|}\; d^3x \;d^3y~~,
\eeq
an expression which amounts to an equivalent of Penrose's in the limit
where the individual eigenstate correlates/conformations 
have the same gravitational self-energy.

Hence, for two states separated by a distance $\Delta r$ which is greater 
than their own spatial extent, this can be approximated as
\beq
E_{\Delta} \sim \frac{Gm^2}{\Delta r}
\label{lifetime}
\eeq
The collapse time of the spacetime superposition is determined by
the uncertainty relation \cite{diosi,penrose1}
\beq
T_c \sim \frac{\hbar}{E_\Delta}~~, 
\label{tcpen}
\eeq
and thus is inversely proportional to the gravitational energy of the system
(calculated in either Di\'{o}si's original or Penrose's alternate 
formalism).  In effect, the
value $T_c$ can be thought of as a ``decay time'' or half-life for the
unstable superposition.

Penrose has used the relations in Equations~(\ref{lifetime}) and (\ref{tcpen})
 to determine
the ``collapse time'' for various states ranging from simple quantum states
({\it e.g.} nucleons) to more mesoscopic objects like specks of dust.  For
a nucleon of mass $10^{-27}~$kg whose superpositions are separated by the
strong interaction scale of $10^{-15}$~m, it can be shown that $T_c \sim
10^{15}~$seconds (or about $10^7$~years).  Hence, this object
will remain superposed for a practically indefinite amount of time, insofar
as laboratory experiments are concerned.  This value is completely reasonable,
since neutron diffractions patterns are observed for a range of
energies (see {\it e.g.} Reference~\cite{neutron}).  

Similarly, Penrose estimates the 
decay times for more macroscopic objects such as specks of dust, water 
droplets, and even cats to show that $T_c$ rapidly dips to small order
of magnitudes, providing again a ``match'' for observation ({\it i.e.}
that macroscopic objects effectively do not maintain measurable
superpositions).

\section{Large extra dimensions and TeV-scale gravity}
In 1998, a revival of Kaluza and Klein's spacetime theory with compactified
dimensions was introduced to explain the hierarchy problem \cite{add}.  
Instead of a single dimension of Planck scale compactification radius, 
the novel framework proposed that there exist $n$ extra flat spatial dimensions
(LEDs) of ``large'' radius $R_n$ into which only gravitation may propagate (no
standard model fields).  It is assumed that there is only one unification
scale -- that of the electroweak scale, $m_{EW} \sim 1~$TeV, and that the absurdly
large gravitational scale in the usual four dimensional space 
$M_{Pl} \sim 10^{16}~$TeV is a perceived artifact of the true nature of
gravity in a higher-dimensional space\footnote{The actual value of
$M_{Pl}$ in the literature varies between $10^{15}~$ and $10^{16}~$TeV.
This effectively results in an order-of-magnitude difference in the scale size
of the extra dimensions, but does not significantly impact the calculations
herein or elsewhere.}

The relationship between the traditional Planck scale and the actual
gravitational unification scale can be shown to be of the form
\beq
M^2_{Pl} \sim R_n^n M^{2+n}_{4+n}~,
\label{scalematch}
\eeq
up to factors depending on the geometries of the manifolds.
Demanding that $M_{4+n} \sim 1~$TeV, the size of the extra 
dimensions is found to be constrained by the relationship
\beq
R_n \sim 10^{\frac{32}{n}-19}~{\rm meters}
\label{radiusn}
\eeq
The relationship in (\ref{scalematch}) can be re-written in terms of the
gravitational couplings in 4- and 4+n dimensions as
(\ref{scalematch}) that
\beq
G_{4+n} \sim G_4 R_n^n~,
\label{g4n}
\eeq
again up to geometric factors.
A more formal derivation of the exact relationship between $G_4$ and
$G_{4+n}$ can be derived any number of ways ({\it e.g.} using Gauss'
law \cite{add}), but for the present analysis the above relationship
will suffice.

The upshot of this new behavior of gravity in the extra dimensions is
a departure of classical Newtonian laws on scales $r < R_n$.  
Instead of the usual $1/r$ potential, the field generated by a mass $m$ 
is now determined by
\beq
|\phi_{4+n}(r)| \sim \frac{G_{4+n}m}{r^{n+1}}
\eeq
and thus the potential energy between two like masses is
\beq
|V_{4+n}(r)| \sim \frac{G_{4+n}m^2}{r^{n+1}}
\label{v4n}
\eeq
again assuming $r < R_n$.  

The possibility of such TeV-scale gravity is a very intriguing one at
present, since future accelerators such as the LHC or Tevatron will be
able to probe precisely these collision energies.  Also, curious behavior
in the cosmic ray flux spectrum shows a distinctive ``knee'' around
energies of 1~TeV, suggesting that some variety of unexplained mechanism
is at work in this region \cite{cr1,cr2}.  
Current table-top gravity experiments have
ruled out extra dimensions (that is, deviations from the inverse-square
law) to a scale of about 200~$\mu$m \cite{eotwash},
which would correspond to between $n=2$ and $3$ in Equation~(\ref{radiusn}).

%Additional constraints on $n$ and/or $R_n$ can be obtained by astrophysical
%measurements, cosmic ray fluxes, and high-energy accelerator experiments,
%although for the time being it will be assumed that extra dimensions on the
%sub-nanometer scales are possible.

\section{OR in a TeV gravity framework}
If the spatial separation of the superposed mass states is small, then
due to the explicit gravity dependence of Equation~(\ref{lifetime})
the existence of LEDs would clearly affect the collapse time\footnote{That
large extra dimensions should affect gravity-driven decoherence mechanisms
such as those discussed herein was originally suggested in \cite{tamas}.
However, this was anecdotal and no calculations accompanied the idea.}.  The 
meaning of the integral in Equation~(\ref{pen1}) remains unchanged, and only
the functional form of the potential changes.  Thus, it can be surmised that

\beq
\Delta E_{G,n} \sim \frac{G_{4+n}m^2}{(\Delta r)^{n+1}} 
\label{egn}
\eeq
which as in \cite{penrose1} can give an order-of-magnitude estimate for
the superposition lifetime, analogous to Equation~(\ref{tcpen}).

%Two distinct
%values of $G_{4+n}$ are used to illustrate a mild discrepancy in the literature.
%The column $T_{c}$ corresponds to calculations usinig a gravitational
%coupling of the form (\ref{g4n}), while $T_{c,\rm approx}$ uses the simpler
%relationship in (\ref{g4napprox}).  While each approximation gives
%%similar values for low $n$, the order of magnitude begins to diverge 
%significantly as $n$ grows.  

Table~\ref{ledcollapse} shows estimated collapse times $T_c$ for the
OR mechanism, assuming the existence of ADD LEDs.  The data is calculated
assuming a gravitational unification scale $M_{4+n}\sim 1~$TeV.  A discussion
of alternate values of $M_{4+n}$ follows at the end of this section,
with the rationale being to place bounds on possible
deviations from the collapse time predictions in Reference~\cite{penrose1}.
As noted in \cite{add}, it is most the value of $M_{4+n}$ which will
be observed as marking the transition to new gravitational interactions
(and not via measurements of $R_n$).

The general conclusion which can be drawn from these figures, however, 
is multifold.  First, if one assumes that Penrose's mechanism is correct,
it places heavy constraints on the possible size of extra dimensions.  If
nucleon superpositions have exceedingly short lifetimes due to the 
much stronger nature of ADD gravity below large $R_n$ ({\it e.g.} for
$n=2$ and $n=3$), then it would be virtually impossible to observe 
coherent neutron diffraction patterns.  The cases $n=4$ and $n=5$ are
interesting, because certainly such short times could be easily observed in
a laboratory setting.  

Since they have not, it is likely that there must
be $n > 5$ extra dimensions with $R_n \leq 10^{-12}~$m.  Furthermore,
as previously mentioned direct measurements of sub-millimeter gravity ($1/r^2$ deviations)
have thus far placed a cap of 200~$\mu$m on the possible size of extra 
dimensions \cite{eotwash} (ruling out $n=4$), 
thus this limit is well below these bounds.  
Alternatively, if the superposition
lifetimes are on the order of a few seconds to a few years ($n=5$ through
$n=7$), it might be possible to devise an experiment to observe this
collapse.  A more recent result has reduced this limit to
100~nm \cite{decca}.

%Although the values of $T_c$ are calculated with the ``correct'' version
%of $G_{4+n}$, those of $T_{c,\rm approx}$ more smoothly transition to 
%the usual Newtonian value obtained by Penrose.  Basically, this distinction
%arises from the extra factors of $2\pi$ in the bulk volume expression
%${\cal V}_n = (2\pi R_n)^n$.  

As the number of dimensions increases beyond $n=7$, 
the scale size becomes
on the order of the supposed state separation of $10^{-15}~$m, and thus
``regular'' Newtonian gravitation will take over.  
The range $n \geq 6$ is 
a particularly interesting regime, since the physical size of the dimensions
shrinks very slowly at this point.  Instead of decreasing by order-of-magnitude
jumps for each $n$, the scale size ``lingers'' in the femtometer region.
Indeed, as $n \rightarrow \infty$, it can easily be seen that $R_n \rightarrow
10^{-19}~$m.

\subsection{Yukawa type gravitational potential}
It should be acknowledged that as the compactification scale 
becomes on the order of the state separation (or lower), the simple 
approximation begins to break down,
and thus the figures may not be completely valid.  In fact, one would
ideally expect the collapse times to return to the original prediction
at a faster rate with an energy function of the form $V_4(r>R_n) + 
V_{4+n}(r<R_n)$, where the former represents the usual Newtonian potential
on the ``brane'', and the latter in all $4+n$ dimensions.

In this spirit, a better approximation for the potential in the cases $n\geq 8$ would be
the Yukawa-type function
\beq
V(r) \sim \frac{G_4m^2}{r}\left(1+2n e^{-r/R_n}\right)
\label{yukawa}
\eeq
for toroidal compactifications, which provides subtle corrections to the 
usual Newtonian potential energy
in the case $r > R_n$ \cite{kehagias}.  This gives $T_c \sim 2\times 
10^{14}~$seconds
for $n=8$, and $5\times 10^{14}~$seconds for $n=9$ -- marginally different
from Penrose's prediction of $T_c \sim 10^{15}~$seconds, but most likely
difficult to measure.

\subsection{Variation of gravitational unification scale}
What if the new Planck scale $M_{4+n}$ is not 1~TeV, but is either slightly
bigger (or even slightly smaller)?  That is, if the scale is $M_{4+n} \sim 
10^\gamma$~TeV (for any integer $\gamma$), 
the scale relationship can be written more generally as
\beq
R_n \sim 10^{\frac{32-2\gamma}{n}-19-\gamma}~{\rm meters}
\label{rnalpha}
\eeq
In this case, it can be shown that the collapse times will be related to
the originals by a factor of $10^{(2+n)\gamma}$.  However, for larger
values of $M_{4+n}$ the femtometer compactification scale will be reached
quicker, thus giving less ``parameter space'' for reasonable collapse times.

Table~\ref{tab2} shows the estimated collapse times for the cases
$\gamma =1$ and $\gamma = 2$ ({\it i.e} $M_{4+n} \sim 10$~TeV and 100~Tev,
respectively).  The transition to regular gravity occurs around
$n=6$ ($\gamma = 1$) and $n=4$ ($\gamma = 2$) extra dimensions, 
with a $T_c \sim 10^{15}$~seconds and $10^{11}~$seconds for each case.
The predictions are still experimentally testable, in principle, since the
collapse times on the order of $10^7~$seconds could be observable in
the experiments discussed in \cite{mirror}.

If the compactification scale is just shy of the TeV range ($\gamma < 0$),
then the Newtonian transition limit occurs at higher $n$.  However, since 
current accelerator experiments probe in the 100~GeV energy range, extra
dimensions signatures would likely have seen as quantum gravity effects 
or missing energy in such beam events.

\subsection{States with $\Delta r > R_n$}
For the larger ``states'' considered by Penrose, the existence of extra
dimensions would barely affect the collapse time since the scale size
drops quickly for small $n$.
A drop of water 100~nm whose superposed states are separated by its radius 
would only show differing collapse times from Penrose's predictions 
for $n=2$ extra dimensions. This would give a $T_c \sim 0.01$~seconds,
versus 1~hour predicted in \cite{penrose1}.  Reiterating the results reported
in \cite{eotwash}, though, $n=2$ has effectively been ruled out.
Since $n=3$ dimensions are on the nanometer scale, any LED-enhanced OR 
effects would be for objects smaller than this.  

One could again apply the Yukawa potential of Equation~(\ref{yukawa}) for
the values where $\Delta r > R_n$, but in most cases this yields effectively
negligible corrections to the usual Newtonian case.

\subsection{Time variation of the gravitational constant}
The notion of time-dependent compactification radii $R_n(t)$ in the ADD 
framework
could yield some interesting ``early universe'' effects if the OR mechanism is
to be believed (see \cite{marciano} for a general overview, although this
reference precedes the notion of ``large'' extra dimensions by 15 years;
\cite{loren} offers a more ``modern'' perspective on the issue).
A review of the current literature indicates that at most 
$\dot{G}_4(t)/G_4(t) < -10^{-11}\;{\rm yr}^{-1}$ \cite{uzan}.

From Equation~\ref{g4n}, this implies $\dot{R}_n(t) > 0$, since the
unification scale is constant.  That is, the compactification scale is growing.
If the radii were
less than the femtometer range in the past, then there should be no significant
deviations from Newtonian gravity for the cases considered herein (although
the mechanism itself will be affected by the different value of $G_4(t)$ at
earlier times).  However, if the radii $R_n(t)$ are growing, then this 
suggests that superposition collapse could be adversely affected in the far 
future.

If it is assumed gravity has been a TeV-scale phenomenon for the entire
existence of the Universe, then it can be deduced from Equation~(\ref{g4n})
that a time-dependent Newtonian constant would behave as 
$G_4(t) \sim R_n^{-n}$, and thus 
\beq
\frac{\dot{R}_n(t)}{R_n(t)} = -\frac{1}{n}\frac{\dot{G}_4(t)}{G_4(t)}~~.
\label{rgt}
\eeq
The simplest exponential solution of these equations indicates that the 
compactification 
radius has increased by only 1-10\% over the history of the Universe 
($10^{10}~$ years) for the number of dimensions of interest, and depending
on the choice of bound for $\dot{G}_4(t)/G_4(t)$.  As a result, the
collapse times would not be appreciably different from their present 
values.  However, an exponential solution is not realistic, since it
implies that the compactification radius would have some non-zero value
at the beginning of the Universe ($t=0$).

Alternatively, it has been proposed that the gravitational constant might
evolve according to a power law of the form $G(t) \sim t^{-\beta}$, where
the exponent $\beta < 0.1$ \cite{demarque}, which corresponds to the bound
$\dot{G}_4(t)/G_4(t) < -10^{-11}~$yr$^{-1}$.  In this case, the LEDs would
behave as $R_n(t) \sim t^{\beta/n}$, with the requirement that at the
present epoch ($t_p$), $R_n(t_p)$ is equivalent to the value given  by
Equation~\ref{radiusn}.  So, the LEDs expand from a trivial size
at the time of the Big Bang to their present size today, in such a way
that 
\beq
\frac{\dot{R}^n_n(t)}{R^n_n(t)} = \frac{\beta t^{-1}}{n}~~.
\label{powerrn}
\eeq
This shows extremely rapid expansion of the extra dimensions in the early 
Universe.  The dependence on $n$ also implies that a smaller number of LEDs
will grow faster.   Smaller earlier dimensions imply that less quantum 
systems would be influenced by the Penrose-Di\'osi mechanism, but that
more systems in the future will come under its influence.

\begin{figure}[h]
\leavevmode \begin{center}
\includegraphics[scale=0.6]{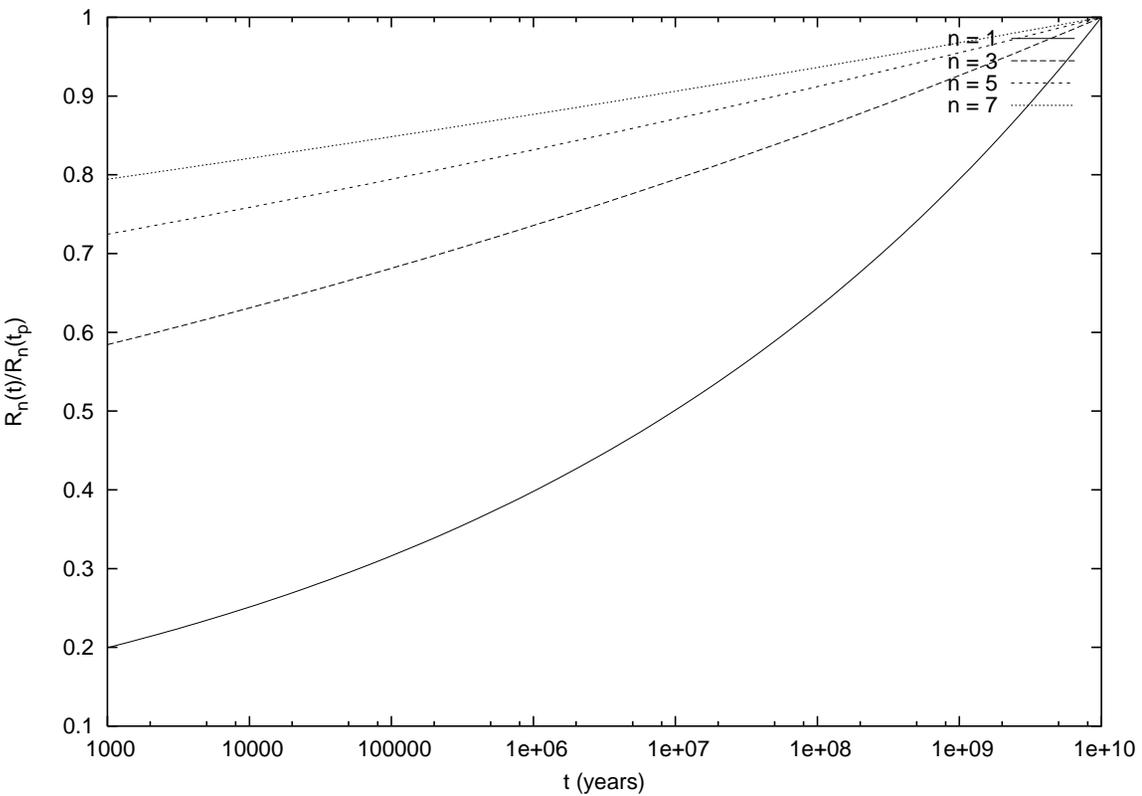}
\caption{Relative variation of compactification radius $R_n(t)/R_n(t_p)$ from
$t=1000~$years after the Big Bang to the present epoch ($t_p = 10^{10}~$years,
assuming a power law dependence on the gravitational constant,
$G_4(t) \sim t^{\beta}, \beta = -0.1$.}
\label{fig1} \end{center}
\end{figure}

Figure~\ref{fig1} shows the relative variation in LED size as
a function of time ($R_n(t)/R_n(t_p)$), from 1000~years after the Big Bang 
until the present epoch, assuming $\dot{G}(t)/G(t) = -10^{-11}$.  
In this prescription, a single LED would have
been only 20\% its present size at this very early age, but a larger
number of LEDs would only have grown a minimal amount.

The changing size of the compactification radius does not change
the collapse time {\it within} that scale, $\Delta r < R_n$, since in that 
region the gravitational coupling is still $G_{4+n}$.  
Since the LEDs would have been even smaller further in the past, this
indicates that most early Universe dynamics would have been 
governed by the rules of quantum mechanics in the absence of extra dimensions.
However, the collapse times will be different because {\it regular} gravity
is stronger than in the present epoch.  This can be see from the fact that 
$\dot{T}_c(t)/T_c(t) = -\dot{G}(t)/G(t)$,
since $T_c(t) \sim 1/G(t)$ from Equation~(\ref{tcpen}).  So, a decreasing
value of $G(t)$ correlates to an increasing value of $T_c(t)$.  In this
case, if $G(t) \sim t^{-\beta}$, then $T_c(t) \sim t^{\beta}$.
When $R_n < 10^{-15}~$m at
an earlier epoch, the nucleon collapse time will be shorter than its 
present value by a factor of $\left(t/t_p\right)^\beta$.

It is also a fascinating consequence that if the extra dimensions continue to
grow, gravity-driven collapse will become more important for macroscopic
systems far in the future.  So, since collapse times might have been shorter
in the past due to stronger regular gravity, and will be shorter again in 
the future due to larger $R_n$, there must be a {\it maximum} value for
the collapse which will be achieved at some point in the future.
It should finally be noted that variation of the gravitational constant 
impacts {\it any} gravity-driven
collapse mechanism ({\it e.g.} see \cite{newdiosi}), 
irrespective of the existence of extra dimensions.

\section{Conclusions and future directions}
The rough calculations presented herein suggest that observable
deviations in quantum superposition behavior could potentially be observed
in a world with extra compactified dimensions.  This implies either serious
problems for the objective reduction mechanism, or alternatively strict 
constraints on the possible size of extra dimensions.  Should extra dimensions
exist, it is apparent that due to the stronger nature of gravity at scales
$r \ll R_n$, the collapse time of quantum superpositions becomes exceedingly
small, and thus in such a scenario the ``size'' of the quantum regime
is also limited.  This should have a major impact on any observational 
prediction which
stems from such a model, including the mirror reflection experiment proposed
in Reference~\cite{mirror}.  

The neutrino oscillation mechanism discussed in \cite{neutrino} lends itself
nicely as a potential test-bed for the OR mechanism with extra dimensions, 
since the interaction scale of the neutrinos should be even smaller than
the femtometer strong interactions considered herein (albeit the much smaller
rest mass will scale the effects accordingly).

Furthermore, a novel application of the 
gravitationally-induced collapse mechanism can be found in 
References~\cite{orchor1,orchor2,orchor3}, which posits that 
``conscious instances'' are merely orchestrated reductions of quantum 
superpositions of entangled microtubulin protein conformation states.  
If the interaction distances are at or below the compactification scale of 
any extra dimensions, this ``Orch-OR'' could greatly impact the theory
and potentially invalidate it \cite{jrmorchor}.

Finally, there is no reason to believe that ADD is the correct theory
of extra dimensions.  The most popular competing theory -- that of 
Randall and Sundrum \cite{rs} (RS) -- may well have its own poignant implications
for superposition collapse in the OR framework, and all the observational
phenomenon which result.  The expected deviations from non-relativistic
Newtonian gravity in the RS framework is of the form
$$V(r) = \frac{G_4 m^2}{r} \left(1+\frac{1}{k^2r^2}\right)$$
where $k$ is the graviton mode, presumed to be on the order of the Planck
scale.  These corrections would be extremely small even for $r\sim 10^{-15}~$m,
and thus may not result in any ``observable'' signature in the Penrose scheme.

\vskip 1cm
\noindent{\bf Acknowledgments}\\
I would like to thank Joy Christian for some insightful discussions on the
philosophical foundations of the mechanism.  I also thank Tamas Geszti
and Lajos Di\'{o}si for pointing out specific literature references and
clarifications on previous results.

\pagebreak

\begin{table}[h]
\begin{center}
\begin{tabular}{c|ccc}\hline
$n$ & $R_{n}$~(m) & $T_{c}$~(s)  \\ \hline 
2 & $10^{-3}$& $10^{-9}$\\
3 & $10^{-9}$& $10^{-5}$\\
4 & $10^{-11}$& $10^{-2}$ \\
5 & $10^{-13}$& $10^1$\\
6 & $10^{-14}$&$10^7$ \\
7 & $4\times 10^{-15}$& $10^{11}$\\ 
8 & $1\times 10^{-15}$& $10^{15}$\\ \hline
\end{tabular}
\caption{Gravity-driven collapse times for nucleon superposition with
unification scale $M_{4+n} \sim 1~$TeV,  assuming
a hard spherical model with spatial separation $\Delta r \sim 10^{-15}~$m.
The collapse time approximation approaches the regular Newtonian value
for $n=8$ since $\Delta r \sim R_n$.}
\label{ledcollapse}
\end{center}
\end{table}

\begin{table}[h]
\begin{center}
\begin{tabular}{c|cc||cc}\hline
 & \multicolumn{2}{|c}{$\gamma=1$} & \multicolumn{2}{||c}{$\gamma=2$}\\ \hline 
$n$ & $R_{n}$~(m) & $T_{c}$~(s) & $R_n$~(m) & $T_c$~(s)  \\ \hline
2 & $10^{-5}$ & $10^{-5}$ & $10^{-7}$& $10^{-1}$\\
3 & $10^{-10}$ & $1$ & $10^{-12}$ & $10^5$\\
4 & $10^{-13}$ & $10^5$ & $10^{-14}$& $10^{11}$\\
5 & $10^{-14}$ & $10^9$ &  --- & ---\\
6 & $10^{-15}$ & $10^{15}$& --- & ---\\
\end{tabular}
\caption{Gravity-driven collapse times for nucleon superposition with
unification scale $M_{4+n} \sim 10~$TeV and $M_{4+n} \sim 100~$TeV,  assuming
a hard spherical model with spatial separation $\Delta r \sim 10^{-15}~$m.}
\label{tab2}
\end{center}
\end{table}

\end{document}